\def\cm{cm$^{-1}$}

\documentclass[prb,twocolumn,showpacs]{revtex4}
\usepackage[dvips]{graphics}
\usepackage{textcomp}
\usepackage{latexsym,cmmib57}
\usepackage{amsmath}
\usepackage{amsthm}
\usepackage[psamsfonts]{amssymb,amsfonts,eucal}

\begin{document} %\draft
\title{Temperature and frequency dependent optical properties of ultra-thin Au films}
\author{Tobby Brandt}
\author{Martin H{\"o}vel}
\author{Bruno Gompf}
\author{Martin Dressel}
\affiliation{1.~Physikalisches Institut, Universit{\"a}t
Stuttgart, Pfaffenwaldring 57, 70550 Stuttgart Germany}

\date{\today}
\begin{abstract}
While the optical properties of thin metal films are well understood in
the visible and near-infrared range, little has been done in the mid-
and far-infrared region. Here we investigate ultra-thin gold films
prepared on Si(111)($7\times 7$) in UHV by measuring in the frequency
range between 500~\cm\ and 7000~\cm\ and for temperatures between 300~K
and 5~K. The nominal thickness of the gold layers ranges between one
monolayer and 9~nm. The frequency and temperature dependences of the
thicker films can be well described by the Drude model of a metal, when
taking into account classical size effects due to surface scattering.
The films below the percolation threshold exhibit a non-metallic
behavior: the reflection increases with frequency and decreases with
temperature. The frequency dependence can partly be described by a
generalized Drude model. The temperature dependence does not follow a
simple activation process.
For monolayers we observe a transition between surface states around 1100~\cm.

\end{abstract}

 \maketitle
%
%\begin{multicols}{2}
%\columnseprule 0pt
%
%
%
%text
\section{Introduction}

The optical properties of thick metal films in the infrared
spectral range are similar to bulk material \cite{Bennett66} and
can be well described by the Drude model, except that interface
scattering becomes increasingly important as the thickness $d$ is
reduced below the mean free path $\ell$, which is approximately
40~nm for the example of gold.\cite{Fahsold00} For extremely thin
films not only the scattering rate increases, but in addition they
exhibit a drop of the effective carrier density, seen in a reduced
plasma frequency;\cite{Pucci06,Walther07} this can be explained by
a modified band-structure and dipole layers at the interfaces.
When the film thickness shrinks even further, normally a
metal-to-insulator transition is observed in the electrical
transport, known as the percolation threshold.\cite{Lai84, Kir73}
The discontinuous film morphology does not only effect the dc
resistivity but also the optical properties. Above the percolation
threshold, the infrared reflectivity decreases with increasing
frequency, it becomes nearly frequency independent at the
threshold, and finally increases with frequency below the
percolation threshold.\cite{Henning99}

During the deposition process, nanometer-sized metal clusters are
created through nucleation and growth. Above a critical
thickness $d_c$, these islands coalesce to form a conducting
network. In principle, one can consider semi-continuous metal films
as composition of metallic particles embedded in an insulating matrix
and try to describe their optical properties with effective
medium theories (EMA).\cite{Max04,Bru35} However, it was shown
that EMA models fail to predict the dielectric behavior of
discontinuous films,\cite{Yagil92, Gom07, Bed01} mainly because
they neglect interface effects.

The temperature dependent dc resistivity of thin metal films above
the percolation threshold contains contributions from regular
phonon scattering, but in addition from grain boundary and surface
scattering. Both processes lead to a thickness-dependent shift of
the (temperature-dependent) resistivity to higher values as the
films get thinner. The overall behavior, however, does not change:
at higher temperatures $\rho(T)$ exhibits a linear dependence,
while at lower temperatures a temperature-independent residual
resistance remains.\cite{deVries87, Mar04} For discontinuous metal
films (below the percolation threshold) Hill found an activated dc
transport, with an activation energy of less than 100~meV for gold
films on glass.\cite{Hill69}

To the best of our knowledge, there exist no temperature dependent
infrared investigations on ultra-thin metal films around the
percolation threshold. In order to improve our understanding of
these films, we have carried out low-temperature infrared optical
studies on gold layers with a nominal thickness varying between
one monolayer and 9 nm.

\section{Experiments}
Thin Au films were prepared by an electron-beam heated effusion cell on
clean Si(111)($7\times 7$) surfaces in UHV at a base pressure below
$10^{-10}$~mbar. During evaporation the samples were at room
temperature and the evaporation rate was about 0.1~nm/min. The film
thickness was monitored by a quartz microbalance. Additionally the
clean surface and the monolayers were characterized by low-energy
electron diffraction (LEED). Whereas the clean Si(111) surface exhibits
the well-known ($7\times 7$)-reconstruction, one monolayer of Au on Si
leads to the formation of a well-ordered Si(111)-Au($6\times 6$)
reconstruction.\cite{Lay83}

After preparation, the films were transferred to an optical UHV
cryostat without breaking the vacuum. The cryostat has one ZnSe
window for reflection measurements in nearly normal incidence and
is attached to a Bruker IFS 113v Fourier-transform infrared
spectrometer. The investigations were performed in the mid- and
near-infrared spectral range between 500~\cm\ and 7000~\cm,
employing a nitrogen cooled mercury-cadmium-telluride (MCT)
detector. The spectra were recorded with a resolution of 8~\cm\
and in each case 128 spectra were averaged. The He cold-finger
cryostat allows us to measure at temperatures down to 5~K. As
reference the same sample was used but coated with an in-situ
evaporated thick gold film. In this way an accuracy in the
reflection measurements better than 1\%\ can be achieved.

\section{Results and Analysis}

The results can be divided in three clearly distinguishable
regimes: The continuous films in the thickness range above 4~nm,
the films close to the percolation threshold around 2~nm, and the
monolayer range.

\subsection{Thick films above the percolation threshold}
\begin{figure}
\scalebox{1.3}{\includegraphics*{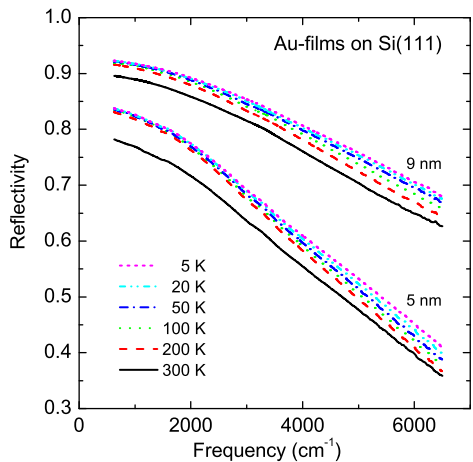}}
\caption{\label{fig:spectrum5_9}(color online) Frequency dependent reflectivity of
thin Au films above the percolation threshold measured at various
temperatures as indicated. For the 9~nm and the 5~nm thick film
the optical behavior can be fitted by a Drude model when classical
size effects are considered.}
\end{figure}
Figure~\ref{fig:spectrum5_9} shows the temperature dependent
reflectivity spectra for a 9~nm and a 5~nm thick film. The
reflectivity decreases with increasing frequency and becomes
better as the temperature is lowered; this behavior corresponds to
normal metals and can be fitted by the Drude
model,\cite{DresselGruner02} in which the real part of the
conductivity is given by:
\begin{equation}
\sigma_1(\omega)=\frac{\omega_p^2\tau}{4\pi}\frac{1}{1+\omega^2\tau^2}\
; \label{eq:drude}
\end{equation}
here $\omega_p$ is the plasma frequency, and $\tau$ denotes the
scattering time. This fit yields the plasma frequency $\omega_p$ and
scattering
 rate $\gamma=1/(2\pi c \tau)$ as listed in Tab.~\ref{tab:drude}.

\begin{figure}
\scalebox{1.3}{\includegraphics*{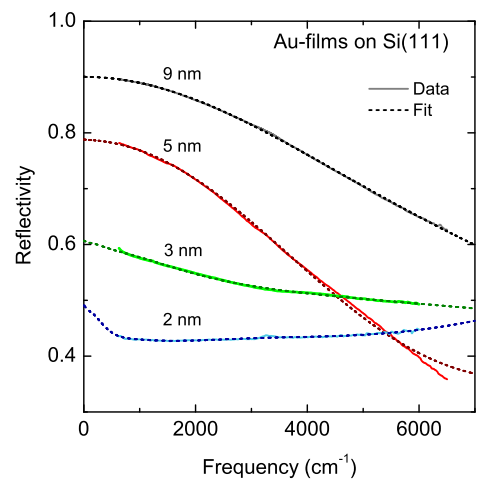}}
\caption{\label{fig:fit}(color online) Frequency dependent
reflectivity of thin Au-films on Si(111) at 300~K together with
the corresponding fits. The 9~nm and the 5~nm films can be fitted
purely by the Drude model with the parameters listed in
Tab.~\ref{tab:drude}. For the 3~nm and the 2~nm film an additional
Lorentz oscillator at higher frequencies is needed for reasonable
description.}
\end{figure}

 \begin{figure}
\scalebox{1.3}{\includegraphics*{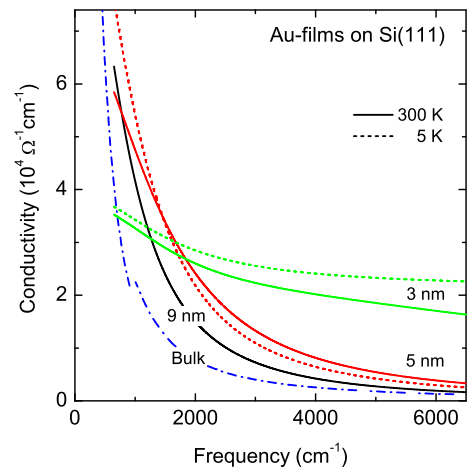}}
\caption{\label{fig:cond3_5_9}(color online) Optical conductivity of
thin gold films on Si(111) obtained from reflectivity measurements on
films with a thickness of $d=3$~nm, 5~nm, and 9~nm, as indicated.
The solid lines correspond to the room-temperature data, the dotted
lines to $T=5$~K. In addition the bulk data (dashed dotted line) are
shown as obtained from Ref.~\protect\onlinecite{Bennett66,DresselGruner02}.}
\end{figure}

\begin{table*}[tb]
\caption{\label{tab:drude} Room temperature properties of thin
metal films obtained from a Drude fit of the optical data.
$\sigma_1(\omega\rightarrow 0)$ corresponds to the dc
extrapolation of the optical conductivity, $\omega_p$ is the
plasma frequency and $\gamma=1/(2\pi c \tau)$ denotes the
scattering rate. For the accurate description of the 3 and 2~nm
films additionally a Lorentz oscillator is needed with the center
frequency $\omega_0$ and the width $\gamma$.}
\begin{ruledtabular}
\begin{tabular}{ccccccccc}
              &\quad &               \multicolumn{3}{c}{Drude components}                &\qquad &    \multicolumn{3}{c}{Lorentz oscillator}      \\
film thickness&      & $\sigma_1(\omega\rightarrow 0)$        & $\omega_p/2\pi c$ & $\gamma$ &       & $\omega_0/2\pi c$ &$\omega_p/2\pi c$ &$\gamma$ \\
(nm)          &      & ($10^4~\Omega^{-1}$\cm)           & ($10^4$~\cm)      & (\cm)     &       & (\cm)             & ($10^4$~\cm)     &(\cm)    \\
\hline
bulk&& 40.9  & 7.27  & 215 &&      &       &      \\
9   && 10.2  & 7.13  & 831 &&      &       &      \\
5   &&  7.0  & 7.81  & 1448&&      &       &      \\
3   &&  3.8  & 6.19  & 1696&& 5380 & 13.21 & 18670\\
2   &&  1.9  & 4.95  & 2165&& 4180 &  7.41 &  8240
\end{tabular}
\end{ruledtabular}
\end{table*}
%
%           |--------------------DRUDE --------------------------|--------------------LORENTZ--------------------|
%film [nm]  sigma_dc [1/(Ohm cm)]    nu_p [1/cm]   gamma [1/cm]      nu_0 [1/cm]     nu_p [1/cm]      gamma[1/cm]
%bulk       408671.39               72718.727       215.808
%9          102105.62369            71320.98803     830.87356
%5           70267.39509            78119.73297    1448.49682
%3           37693.66164            61911.88619    1696.01482        5381.34014      132131.91938     18668.94067
%2           18899.17557            49530.507456   2164.973548       4178.70583       74140.83439      8243.76371
%All values at 300K
%Drude fit to
%[Bennett, H.E. and Bennett, J.M., booktitle: Optical Properties and Electronic Structure of Metals and Alloys, title: Validity of the Drude Theory for %Silver, Gold and Aluminum in the Infrared, 1966, editor: Abelès, F.]
%results:
%sigma_dc= 408671.39  1/(Ohm cm)
%nu_p=      72718.727 1/cm
%gamma=       215.808 1/cm

In Fig.~\ref{fig:fit} a comparison between the 300 K measurements
and the calculated reflectivity for the layer system: fitted
Drude-metal/Si is additionally shown. Included in
Fig.~\ref{fig:fit} as well as in Tab.~\ref{tab:drude} are also the
Drude parameters for the 3~nm and the 2~nm film, although these
films cannot be described by a pure Drude model. For there
description an additional Lorentz oscillator\cite{DresselGruner02} at higher frequencies
is needed:
\begin{equation}
\sigma_1(\omega)=\frac{\omega_p^2\tau}{4\pi}
\frac{\omega^2/\tau^2}{(\omega_0^2-\omega^2)^2+\omega^2/\tau^2}\
; \label{eq:lorentz}
\end{equation}
with $\omega_0$ the center frequency.
 The broad oscillator at higher frequencies reduces the
plasma frequency in the Drude term, which together with the higher
scattering rate leads to the lower reflectivity at lower
frequencies for these films (see Fig.~\ref{fig:fit}). The 5~nm and
9~nm layer exhibit plasma frequencies, which are close to that of
bulk Au.\cite{Bennett66} As for bulk metals, the plasma frequency
in all films is temperature independent. The higher scattering
rate for thinner films is a typical size effect: thinner films
exhibit a stronger surface scattering due to a larger
surface-to-volume fraction.

\begin{figure}
\scalebox{1.3}{\includegraphics*{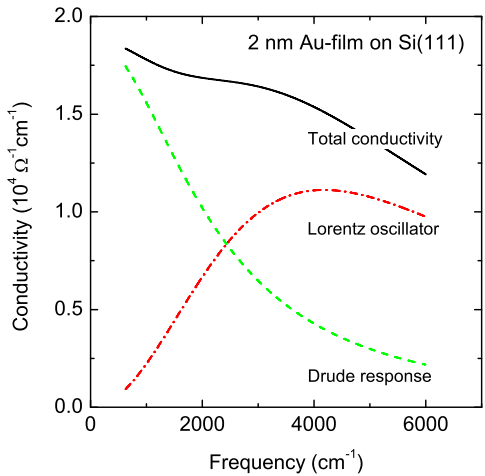}}
\caption{\label{fig:con2nm}(color online) Total optical
conductivity of a 2~nm thick Au film on Si at 300~K together with
the two contributions: the Drude response of the free electrons at
low frequencies and a broad Lorentz oscillator at higher
frequencies.}
\end{figure}

With the Eqs.~(\ref{eq:drude}) and (\ref{eq:lorentz}) and the parameters listed in
Tab.~\ref{tab:drude} the corresponding conductivity of the films
can be calculated (Fig.~\ref{fig:cond3_5_9}). By comparison of the
5~nm with the 9~nm film, which both can be described by a pure
Drude, it becomes obvious that an increase in scattering rate for
thin films leads to an enhanced conductivity in the mid-infrared
range. For the two thinner films the strongly enhanced
mid-infrared conductivity is mainly due to the Lorentz oscillator [Eq.~(\ref{eq:lorentz})]
as can be seen in Fig.~\ref{fig:con2nm}, where the different
contributions to the conductivity for the 2~nm thick film are
shown separately.
\begin{figure}
\scalebox{1.3}{\includegraphics*{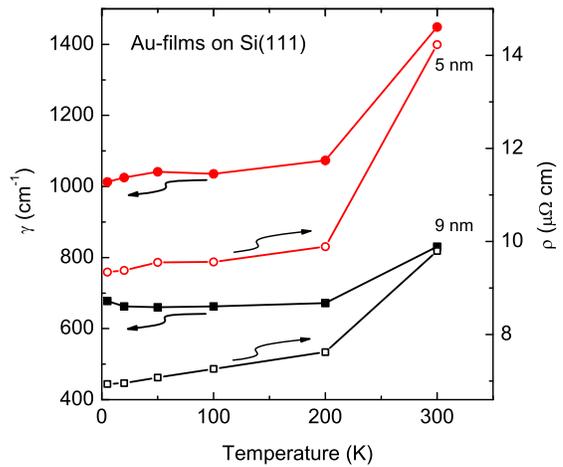}}
\caption{\label{fig:streurate}(color online) Temperature dependence of
the scattering rate $\gamma=1/(2\pi c\tau)$ and resistivity
$\rho=1/\sigma_1(\omega\rightarrow 0)$ for the 9~nm and 5~nm gold
film. The full symbols correspond to the scattering rate (left scale),
the open symbols correspond to the resistivity (right scale).}
\end{figure}

From the extrapolation $\sigma_1(\omega\rightarrow 0)$ we can
estimate the dc resistivity. As visualized in
Fig.~\ref{fig:streurate}, the scattering rate and resistivity of
the 9~nm and the 5~nm film drop strongly as the temperature is
reduced, but the drop is considerably enhanced for the 5~nm film.
That is an unexpected result; scattering on topological rough
interfaces should be temperature independent. The large influence
of temperature on the scattering rate observed for thinner films
is an indication that surface phonons play an important role in
surface scattering. For $T<200$~K phonons are frozen out and the
scattering rate and resistivity remain almost constant with $T$
for both samples.

\subsection{Thin films at the percolation threshold}

\begin{figure}
\scalebox{1.3}{\includegraphics*{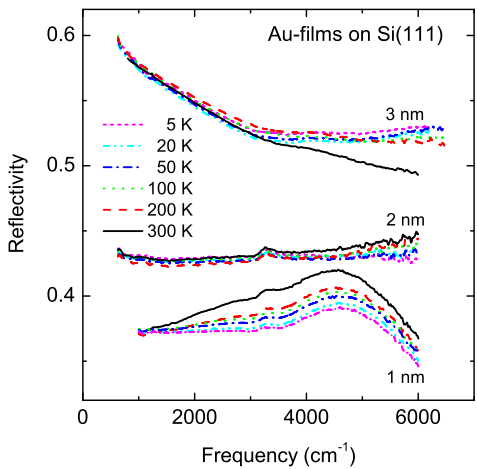}}
\caption{\label{fig:spectrum123}(color online) Frequency dependent
reflectivity of three Au films (1~nm, 2~nm, and 3~nm) at the percolation
threshold. Note the enlarged scale compared to
Fig.~\protect\ref{fig:spectrum5_9}.}
\end{figure}
At a thickness of approximately 2~nm, the Au films undergo a
metal-to-insulator transition which becomes already obvious by
looking at the frequency and temperature independent reflectivity
(Fig.~\ref{fig:spectrum123}). For the 1~nm film the reflectivity
below 4500~\cm\ changes its slope indicating a vanishing metallic
behavior. At low frequencies the metallic contribution of the 3~nm
and 2~nm film can still be analyzed by the Drude model with
basically temperature independent parameters. As summarized in
Tab.~\ref{tab:drude} the values are in accord with the tendency
observed for the 9~nm and 5~nm film: the scattering is even more
dominated by grain boundaries, imperfections and surface roughness
but additionally the plasma frequency starts to drop for the
thinner films. Around the percolation threshold not only the
frequency dependence changes its slope, but also the temperature
dependence is reversed. Whereas the 3~nm film especially in the
near-infrared still shows a slight increase as the temperature is
lowered down to 5~K, a tiny decrease is observed for the 2~nm
film. This is much more pronounced for the 1~nm film, which shows
clear indication for activated transport. All three films show in
addition to their Drude component a contribution of a Lorentzian
line. For the 2~nm thick film this Lorentzian oscillator at
4180~\cm\ makes the reflectivity nearly frequency independent. For
the 1~nm film this oscillator is shifted to slightly higher
frequencies and even more pronounced.

The most remarkable point is that for higher frequencies, i.e.\ in
the near-infrared, the optical reflectivity of films at the
percolation threshold exceed the one of thick films. Although it
seems counter-intuitive, it is a simple result of the fact that
the reflectivity of a metal continuously drops with frequency,
while it remains constant or even increases for films at and below
the percolation threshold. There is an additional dielectric
contribution to the electrodynamic response due to polarization
effects: the isolated clusters interact capacitively.

\subsection{Ultra-thin films}

\begin{figure}
\scalebox{1.3}{\includegraphics*{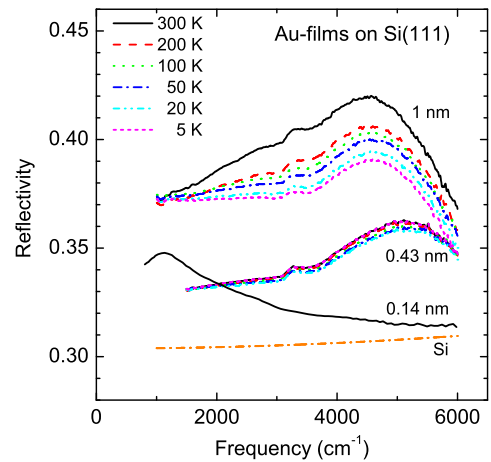}}
\caption{\label{fig:spectrum1043014}(color online) Reflectivity for ultra-thin
films below the percolation threshold. Here the temperature
dependence is reversed compared to metallic transport. The dash
dotted line corresponds to the reflectivity of bare silicon.}
\end{figure}

Below the percolation threshold, i.e.\ for a thickness $d\leq
2$~nm, the temperature and frequency characteristics are inverted
compared to a metal, as shown in Fig.~\ref{fig:spectrum1043014}. In
addition, the reflectivity of bare silicon is displayed in the
figure. For the monolayer corresponding to a nominal thickness of
0.14~nm the optical reflectivity approaches the one of bare Si,
except a strong excitation observed around 1100~\cm\ which we
ascribe to localized surface excitations.

\begin{figure}
\scalebox{1.3}{\includegraphics*{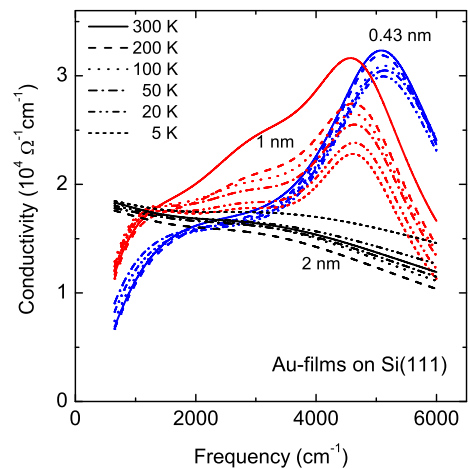}}
\caption{\label{fig:cond2_1_043}(color online) Conductivity spectra of
ultra-thin Au films as derived from reflectivity measurements (Figs. \protect\ref{fig:spectrum123} and \protect\ref{fig:spectrum1043014}) with a nominal thickness of 2~nm, 1~nm, and 0.43~nm.}
\end{figure}

From our temperature dependent optical experiments we can get
information about the high-frequency transport mechanism of
ultra-thin gold films. In the near-infrared range the optical
conductivity increases as the coverage is lowered (see
Fig.~\ref{fig:cond2_1_043}). A strong and broad resonance like
feature around 4700~\cm\ dominates the conductivity for the 1~nm
film. For the 2~nm and even for the 3~nm film the onset of a yet
overdamped resonance at lower frequencies can be identified. This
feature becomes more dominant and sharper as the metal clusters
get further separated and smaller. It cannot be ascribed to
plasmon excitations of individual metal clusters commonly found in
the visible spectral range.\cite{Link99,Scaffardi05} It is most
likely due to a ``Maxwell-Garnett resonance'', which is known to
shift with the area fraction of the substrate covered by
metal.\cite{Dor66,Mar71} Due to the high refractive index of our
substrate this resonances are in the case of Si expected to be in
the infrared.

\begin{figure}
\scalebox{1.3}{\includegraphics*{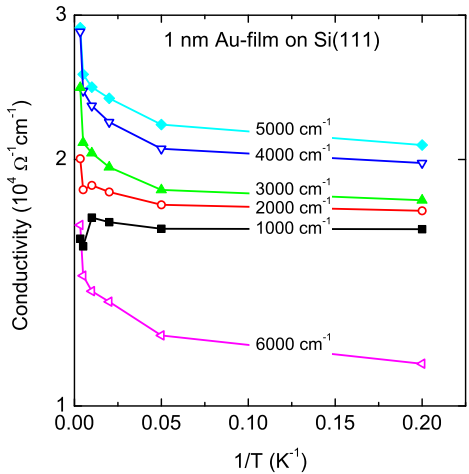}}
\caption{\label{fig:cond1}(color online) Arrhenius plot of the
temperature dependence of the optical conductivity of a 1~nm
Au film at different frequencies as indicated. No clear
indication of a simple activated transport can be found.}
\end{figure}

When the temperature is reduced, the peak does not shift but the
overall-intensity of the infrared conductivity decreases. In
Fig.~\ref{fig:cond1} the logarithm of the optical conductivity at
various frequencies is plotted as a function of inverse
temperature. In this Arrhenius plot, no straight line can be
extracted over a large temperature range, and thus we cannot
identify a simple activated transport behavior from our optical data
as suggested by Hill\cite{Hill69} for the dc conductivity.

\section{Discussion}
The dc resistivity of thin gold films has been investigated as a
function of grain diameter and temperature by various
groups.\cite{deVries87,Chen05} The granularity is varied by
annealing the films and using different substrate materials;
typically the film thickness is well above 10~nm. The importance
of grain-boundary scattering as suggested by the model of Mayadas
and Shatzkes \cite{Mayadas70} can explain most of the findings: a
decrease of resistivity with increasing thickness and grain size,
and the same linear temperature dependence above $T=50$~K.
However, our data displayed in Fig.~\ref{fig:streurate} obtained
from the optical conductivity show a somewhat different behavior.
Below 200~K the scattering rate and resistivity of the 5~nm and
9~nm film exhibit a similar but weaker temperature dependence
compared to $T>200$~K. In the high temperature range the 5~nm film
exhibit a much steeper increase as $T$ rises compared to the
thicker films. Additional processes like surface phonons have to
be responsible for this behavior. Besides the Au surface, the
interface between silicon and gold has to be taken into account.
This interface was studied by a variety of methods which all came
to the conclusion that there is some intermixing due to the high
solubility of Au in Si.\cite{Lay83} This few monolayer thick
intermixing region may become more important as the film thickness
decreases.

We can fit the dominant mid-infrared peak (cf.\
Fig.~\ref{fig:cond2_1_043}) by a simple Lorentzian peak [Eq.~(\ref{eq:lorentz})], centered
around $\omega_0/2\pi c =4180$~\cm, 4733~\cm, and 5147~\cm\ for $d=2$~nm, 1~nm, and
0.43~nm, respectively. It gets obvious that with decreasing film
thickness the peak shifts to higher frequencies.
%subtraction of the Lorentzian line
%film      Temp[K]    nu_0 [cm^-1]    nu_p [cm^-1]      gamma [cm^-1]
%2nm       300        4178.70583      74140.83439       8243.76371
%1nm          300        4733.40604      49257.03191       2149.09529
%1nm            5        4733.31556      41521.41168       1978.96089
%0.43nm    300        5147.53783      56692.92455       2413.84409
%0.14nm    300        there has been no subtraction of a Lorentzian line
%  2008-08-12 calculated by Martin Hoevel
%
In a simple picture, these oscillatory contributions are ascribed to
the capacitive coupling of the metallic clusters. Efros and Shklovskii
considered such a situation and analyzed the critical behavior of
conductivity and dielectric constant near the metal-insulator
transition.\cite{Efros76} They predicted a divergence of the dielectric
constant at a certain filling fraction, which happens distinctively
below the percolation threshold. Since the particle size and coverage
is statistically distributed and interaction with the Si substrate has
to be taken into account, a quantitative analysis of our data is not
really possible in this regard.

\begin{figure}
\scalebox{1.3}{\includegraphics*{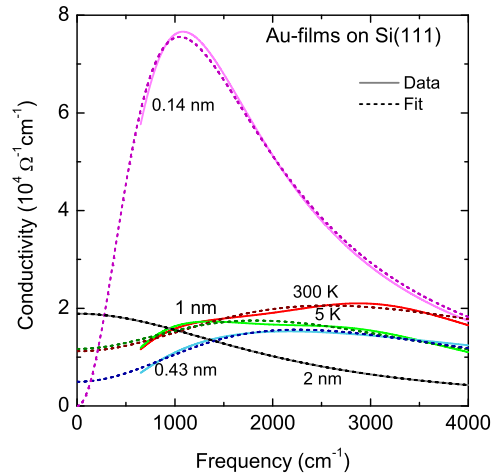}}
\caption{\label{fig:smith}(color online) Fit of the low-frequency
conductivity spectra by Eq.~(\protect\ref{eq:smith}). The Lorentz
contribution was subtracted to focus on the itinerant electrons.
The dashed lines are the experimental data taken from
Fig.~\protect\ref{fig:cond2_1_043}; the solid lines correspond to the
generalized Drude model which also takes backscattering into
account.}
\end{figure}

Instead we want to consider the optical conductivity which remains
after subtraction of the Lorentzian contribution from the measured
conductivity spectra (Fig.~\ref{fig:smith}). For thick films this
should be the itinerant charge transport which becomes gradually
localized as the metallic particles get increasingly disconnected and
scattering at the boundaries is enhanced. Using the single-scattering
approximation of the generalized Drude formula developed by
Smith,\cite{Smith01} the real part of the conductivity is fitted by
\begin{equation}
\sigma_1(\omega)=\frac{\omega_p^2\tau}{4\pi}\frac{1}{1+\omega^2\tau^2}
\left[1+s\frac{1-\omega^2\tau^2}{1+\omega^2\tau^2}\right]\ . \label{eq:smith}
\end{equation}
Here the second term describes the backscattering leading to some
localization of the charge carriers. Accordingly the conductivity
drops for $\omega\rightarrow 0$ as $s$ decreases from $0$ (regular
Drude behavior) to $-1$ (complete localization with
$\sigma_\text{dc}=0$). Below the percolation threshold
no conduction path persists and the dc conductivity vanishes.
In contrast to localization by impurities (Anderson localization) or
electronic correlations (Mott localization), here we observe a geometrical localization by the formation of islands on a nanometer scale.
Applying
Eq.~(\ref{eq:smith}), we obtain $s=-1$ for the $d=0.14$~nm film.
As the film thickness increases backscattering becomes less
important and the value of $s$ changes to $-0.84$ and $-0.71$ for
$d=0.43$~nm and 1~nm respectively. In absence of the Lorentzian
contribution the 2~nm film shows a Drude like behavior ($s=0$)
with very low effective carrier density and high scattering rate
(cf.~Tab.~\ref{tab:smith}). As one example for the temperature
dependence of the Drude-Smith parameters, we have listed for the
1~nm film also the 5~K values. In the framework of the model, not
only the scattering rate and the plasma frequency become
temperature dependent, but also the $s$-parameter. In addition an
extrapolation of the generalized Drude formula to
$\omega\rightarrow 0$ leads to the strange result that the 1~nm as
well as the 0.43~nm film exhibit a non vanishing dc conductivity
for films below the percolation threshold. In this context it
becomes obvious that the applicability of Eq.~(\ref{eq:smith})
seems to be restricted and that the plasma frequency as well as
the scattering rate lose their normal meaning. For the thinnest
film (0.14~nm), for example, the plasma frequency in the
generalized Drude description is higher than the bulk value.
\begin{table}
\caption{\label{tab:smith} Parameters obtained by fitting the
low-frequency part of the conductivity of thin metal films after
subtracting the oscillatory contribution, as described in the
text. The Drude-Smith formula [Eq.~(\protect\ref{eq:smith})] contains an
additional backscattering term, which is maximal for $s=-1$ and
vanishes for $s=0$. The temperature is denoted by $T$, $\omega_p$
is the plasma frequency, $\gamma=1/(2\pi c \tau)$ the scattering
rate.}
\begin{ruledtabular}
\begin{tabular}{ccccc}
film thickness & $T$ & $\omega_p/2\pi c$ & $\gamma$ & s\\
(nm) & (K) & $(10^4$ \cm) & (\cm) &\\
\hline
2    &  300 &  4.95 &  2165 &  0\\
1    &  5   &  6.95 &  2410 &  -0.65\\
1    &  300 &  8.65 &  3120 &  -0.71\\
0.43 &  300 &  6.79 &  2480 &  -0.84\\
0.14 &  300 &  9.25 &  1050 &  -1
\end{tabular}
\end{ruledtabular}
\end{table}

At a coverage of one monolayer (0.14~nm), Au on Si(111)($7\times
7$) forms a well ordered Si(111)-Au($6\times 6$) reconstruction,
which can be clearly identified by LEED. Beside the absence of the
mid-infrared peak, this well ordered surface shows an excitation
at 1083~\cm, which we attribute to a transition between filled
and empty surface states. This excitation was not identified by
photo-electron spectroscopy until now and demonstrates that
FTIR-spectroscopy is a powerful complimentary technique for the
characterization of metal-monolayers.

\section{Conclusion}
Above the percolation threshold the optical behavior of Au films
in the infrared spectral range can be described by the Drude
model, when classical size effects are considered. At the
percolation threshold the optical response becomes frequency and
temperature independent. Below the threshold the optical
properties of ultra-thin metal films are reversed: the
reflectivity increases with frequency and decreases with
temperature. Whereas the frequency dependence can in principle be
described by a generalized Drude formalism, the temperature
behavior cannot be fitted to a simple Arrhenius-like activated
transport. For monolayers, we could identify infrared transitions
between surface states located around 1100~\cm. To uncover the underlying physical
processes leading to the observed high frequency temperature
dependence of the conductivity further investigations on different
metal-substrate systems around the percolation threshold are in
progress.

\section{Acknowledgement}
We would like to thank N.~Drichko for the valuable help during the
optical experiments.

\end{document}